\documentclass[11pt,onecolumn]{article}

 \usepackage[utf8]{inputenc}

\usepackage{geometry}

\usepackage{algorithmicx,algorithm}

\usepackage{tikz}
\usepackage{pgflibraryarrows}
\usepackage{pgflibrarysnakes}
\usetikzlibrary{automata, positioning}
 
\usepackage{ifpdf}
\usepackage{cite}  

\usepackage{xcolor}
\usepackage{float}

\usepackage[noend]{algpseudocode}
\usepackage{srcltx}
\usepackage{mathrsfs}
\usepackage{amsopn}
\usepackage{latexsym,amssymb}
\usepackage{graphicx,color}
\usepackage{enumerate}
\usepackage{amsmath}
\usepackage{booktabs,balance}
\usepackage{flushend}
  
\usepackage{subfigure}

\DeclareMathOperator{\col}{Col}

\definecolor{LGreen}{rgb}{0.8,0.9,0.8}

\definecolor{LRed}{rgb}{0.9,0.8,0.8}
\definecolor{LBlue}{rgb}{0.6,0.6,0.95}

\definecolor{LGray}{rgb}{0.85,0.85,0.85}

\DeclareMathOperator{\idx}{Idx}

\newtheorem{theorem}{Theorem}
\newtheorem{proposition}{Proposition}
\newtheorem{corollary}{Corollary}
\newtheorem{remark}{Remark}
\newtheorem{lemma}{Lemma}
\newtheorem{definition}{Definition}

\newtheorem{example}{Example}

\newcommand{\ddelta}{{\boldsymbol\delta}}

\begin{document}

\title{\bf Robust Set Stability of Logic Dynamical Systems with respect to Uncertain Switching\thanks{This work was supported by the National Natural Science Foundation of China (61873284).}
\thanks{Corresponding author: Yuqian Guo (gyuqian@csu.edu.cn)}}
\author{Yuqian Guo, Zhitao Li\vspace{0.2cm}\\\vspace{0.1cm}\it  School of Automation, Central South University, Changsha, China, 410083\\\small Emails: gyuqian@csu.edu.cn (Y. Guo), lizhitao@csu.edu.cn (Z. Li)}

\date{}

\maketitle

\begin{abstract}
This paper proposes several definitions of robust stability for logic dynamical systems (LDSs) with uncertain switching, including robust/uniform robust set stability and asymptotical (or infinitely convergent)/finite-time set stability with ratio one. It is proved herein that an LDS is robustly set stable if and only if the destination set contains all loops (i.e., the paths from each state to itself); an LDS is uniformly robustly set stable, or finite-time set stable with ratio one, if and only if all states outside the destination set are unreachable from any self-reachable state; and an LDS is asymptotically set stable with ratio one if and only if the largest robustly invariant subset (LRIS) in the destination set is reachable from any state. In addition, it is proved that uniform robust set stability implies robust set stability, and robust set stability implies asymptotical set stability with ratio one. However, the inverse claims are not generally true. The relations between robust stability and stability under random switching are revealed, that is, the asymptotical/finite-time set stability with ratio one under uncertain switching is equivalent to asymptotical/finite-time set stability of the LDS under random switching. Furthermore, it is proved that, for uniform set stability and asymptotical/finite-time set stability with ratio one, the set stability is equivalent to the stability with respect to the LRIS in the destination set. However, robust set stability does not imply robust stability with respect to the LRIS in the destination set. This finding corrects a result in a previous study.

\end{abstract}

	\section{Introduction}

Logic dynamical systems (LDSs) are discrete-time dynamical systems with finite states consisting of collections of discrete-valued nodes in finite logic domains and collections of logic functions governing the state update of all nodes. Boolean networks (BNs), proposed by Kauffman as a qualitative model for gene regulatory networks (GRNs)\cite{kauffman1969metabolic}, are a special type of LDS wherein each node takes a binary value. Although BN models adopt simple regulation relations between neighboring genes, they are capable of predicting long-term qualitative behaviors of GRNs. Apart from GRNs, LDSs have found applications in many different fields, including chemistry, social networks, economics, and computer science \cite{akutsu1999identification, albert2000dynamics,shmulevich2002probabilistic, shmulevich2002boolean, aldana2003boolean, heidel2003finding, akutsu2007control}. During the last two decades, research on LDSs has attracted renewed attention with the invention of the semi-tensor product (STP) of matrices (which is a generalization of the conventional matrix product) and the proposal of the vector representation of logic \cite{cheng2011analysis}. Both the STP and vector representation of logic have become powerful tools for the analysis and synthesis of LDSs \cite{cheng2011stability, qi2010stabilization,Li2014stability, cheng2009controllability, li2011controllabilityProb, li2012complete, Chen2017Synchronization1, cheng2011disturbance, cheng2011identification, laschov2011maximum, Liu2017Pinning,zhao2011optimal}.

In real-world biological systems, uncertainties and perturbations are extremely common\cite{2007Stochastic,LEE2009927}; however, robustness is also a fundamental feature observed in various biological systems \cite{2004Biological}. Robustness analyses of LDSs, as important models for GRNs, aid in the understanding of biological robustness. To simulate the stochastic (i.e., non-deterministic) nature and uncertainties of GRNs, Shmulevich proposed the probabilistic Boolean network (PBN) model \cite{shmulevich2002probabilistic, shmulevich2002boolean,Shmulevich2010Probabilistic}. PBNs are essentially LDSs with random switching wherein the switching signals are typically assumed to be independent and identically distributed (i.i.d.), subject to certain probability distributions. Using the STP, many PBN analysis and design problems have been studied in recent years, such as controllability and observability\cite{Liu2015Controllability, Li2015Weak}, stability and stabilization \cite{Li2016State, Li2014State,  Li2014Partial, Chen2014Stability, Denic2009Robust, wu2016stochastic}, and synchronization \cite{Lu2015Synchronization, Chen2017Synchronization}.   Disturbances and uncertainties are also handled by modeling them as arbitrary switching and analyzing the robustness of certain LDS properties.
Several studies regarding the robust stability of BNs have recently been published. For instance, in \cite{ARACENA20091}, the robustness of BNs with respect to update schedules was analyzed. More recently, the robustness of the stability of BNs and PBNs with stochastic function perturbations were studied in \cite{2020Robustness}, \cite{Ren2021Robust}, and \cite{Li2022Robust}, respectively, using the STP and vector representation of logic. For BNs subject to uncertainties, robust stability with respect to fixed points and loops were studied in \cite{Li2017Robust} and \cite{2017Global}. Subsequently, the robust stability of delayed BNs with respect to fixed points was studied in \cite{2020Robust}.

Among the multitude of scenarios, stability with respect to the subsets has been our focus. For instance, as reported in \cite{Guo2015Set}, the problems of synchronization and stability with respect to partial state variables can be reformulated as a problem of stability with respect to the subsets.

In this paper, we propose different definitions of robust set stability for LDSs. The first definition is simply referred to as robust set stability. An LDS is robustly stable with respect to the destination set if, for any initial state and any switching signal, a solution converges to the destination set within a finite time. Under this definition, although every solution converges to the destination set within a finite time, the transient periods are not required to be uniformly bounded with respect to all switching signals. This motivates the second definition of robust stability, which is referred to as uniform robust set stability. An LDS is said to exhibit uniform robust stability with respect to the destination set if there are finite time steps within which a solution starting from any initial state and under an arbitrary switching signal converges to the destination set. We prove herein that robust set stability and uniform robust set stability become equivalent under one of the following situations:
 \begin{itemize}
   \item The destination set is a singleton (i.e., has only one element).
   \item The LDS under consideration is uncertainty-free, that is, there is only one subnetwork.
   \item The destination set is a robustly invariant set.
 \end{itemize}
However, in general, although uniform robust set stability implies robust set stability, the inverse implication is false because, for an LDS that is merely robustly set stable, the supremum of the transient periods of all solutions may be unbounded. This implies that the results for the stability of uncertainty-free LDSs and the robust stability of LDSs with respect to a fixed point are not directly generalizable to the robust set stability of uncertain LDSs.

The third and the forth definitions of robust stability are respectively termed asymptotical and finite-time set stability with ratio one. Each switching signal is herein referred to as a switching pattern. For a given initial state and a destination set, a reachable switching pattern is a switching signal under which a solution starting from the initial state would reach the destination set. The percentage of the number of reachable switching patterns is referred to as the reachable pattern ratio. An LDS is said to be asymptotically set stable with ratio one if the reachable pattern ratio tends to one as the time goes to infinity. It is called finite-time set stable with ratio one if the reachable pattern ratio converges to one in a finite time. Finite-time set stability with ratio one and uniform robust set stability are essentially equivalent. In addition, robust set stability implies asymptotical set stability with ratio one; however, the inverse implication does not hold. The two switching pattern ratio-based definitions of robust stability also bridge the robust stability of LDS under uncertain switching and the stability of LDSs under random switching. We prove that asymptotical and finite-time set stability with ratio one are equivalent to finite-time set stability with probability one and asymptotical set stability in distribution of LDSs under i.i.d. random switching.

In this study, a reachability analysis was adopted to obtain the criteria for different robust set stabilities. A destination state is said to be reachable from the initial state if there is a switching signal under which a solution starting from the initial state would reach the destination state in a finite time.  Notably, a state is said to be self-reachable if it is reachable from itself. Reachability and self-reachability can be easily verified using the reachability matrix. Specifically, the reachability matrix is a nonnegative matrix, each element of which indicates the reachability between a pair of initial and destination states. In particular, the self-reachability of each state can be checked based on the non-zeroness of the corresponding diagonal element. In addition, the reachability is equivalent to the existence of a path from the initial state to the destination state in the state transition graph (STG), and the self-reachability of any state is equivalent to whether the state is on a loop; here, the STG is defined as the union graph of all subnetwork STGs.

 The concept of largest invariant subset (LIS) has been applied to set stability analysis for both deterministic logic dynamical systems and probabilistic LDSs in previous studies \cite{Guo2015Set,Guo2019Stability}. The concept of largest robustly invariant subset (LRIS) is introduced herein for robust set stability analysis. A robustly invariant subset is a set of states wherein every solution starting from the subset remains within it indefinitely under arbitrary switching. The union of all robustly invariant subsets in a given set is referred to as the LRIS in the set.

 Based on the concepts of reachability and LRIS, we herein prove the following easily verifiable stability criteria:
 \begin{itemize}
   \item An LDS is robustly set stable if and only if all states outside the destination set are self-unreachable, or equivalently, if and only if the destination set contains all loops in the STG of the LDS.
   \item An LDS is uniformly robustly set stable, or finite-time set stable with ratio one, if and only if all states outside the destination set are unreachable from any self-reachable state.
   \item An LDS is asymptotically set stable with ratio one if and only if the LRIS in the destination set is reachable from any state.
 \end{itemize}

In addition, we present the following LRIS-based criteria for uniform robust set stability and asymptotical set stability with ratio one:
\begin{itemize}
  \item An LDS is uniformly robustly set stable if and only if it is robustly stable with respect to the LRIS in the destination set.
  \item An LDS is asymptotically set stable with ratio one if and only if it is asymptotically stable with ratio one with respect to the LRIS in the destination set.
\end{itemize}
However, there is no similar LRIS-based criterion for robust set stability, that is, an LDS that is robustly stable with respect to the destination set is not necessarily robustly stable with respect to the LRIS in the destination set.
This finding corrects the LRIS-based criterion for the robust set stability of LDSs in \cite{Guo2017Invariant}. Specifically, the statement ``\textit{a BN is set stable under arbitrary switching if and only if it is stable with respect to the LRIS in the destination set}'' in \cite{Guo2017Invariant} is incorrect; this can be corrected by replacing the definition of set stability with the uniform robust set stability proposed in this paper.

The remainder of this paper is structured as follows. In Section \ref{SectionSTP}, some necessary preliminaries on the STP and vector representation of logic are introduced. In Section \ref{SectionDefinitions}, the algebraic form of an LDS is derived, and several definitions of robust set stability are proposed. Sections \ref{SectionRobustStability}, \ref{SectionUniformRobustStability}, and \ref{SectionStabilityRatioOne} investigate robust set stability, uniform robust set stability, and stability with ratio one, respectively. Section \ref{SubSecRelations} discusses the relations between different robust set stabilities, and Section \ref{SectionConcluding} presents some concluding remarks. Finally, Table \ref{Notations} lists the notations used in this study.

 \begin{table}[t]
\caption{Notations}
\tabcolsep3.5pt
\renewcommand\arraystretch{1}
\begin{center}\footnotesize
\begin{tabular}{rl}
\toprule
Notations & Explanations\\
\midrule
$\mathbb Z$& set of integers\\
$\mathbb Z_{\geq a}$& set $\{k\in\mathbb Z\bigm | k\geq a\}$\\
$[n:m]$& set $\{k\in\mathbb Z\bigm | n\leq k\leq m\}$ \\
$\mathscr D_n$ & $n$-valued logic domain $[0:n-1]$\\
$\ltimes$ & semi-tensor product (STP)\\
$\otimes$ & Kronecker product\\
$*$ & Khatri--Rao (K-R) product\\
$\col_j(A)$ & $j$th column of matrix $A$\\
$ \alpha$ & vector form of logic variable $\alpha$ \\
$\ddelta_n^j$ & $j$th column of identity matrix $I_n$ \\
$\Delta_n$ & $n$-valued logic domain in vector form\\
$\idx( \alpha)$ & index of logic variable $ \alpha$ in vector-form\\
$\idx(\mathcal M)$ & index set $\{\idx( \alpha)\bigm|  \alpha\in\mathcal M\}$\\
$\mathscr L_{n\times m}$ & set of $n\times m$ logic matrices\\
$\mathscr B_{n\times m}$ & set of $n\times m$ Boolean matrices\\
$\ddelta_n[i_1,i_2,\cdots i_m]$ & logic matrix $L$ with $\col_j(L)=\ddelta_n^{i_j}$, $j\in[1:m]$ \\
$[A]_{i,j}$ & $(i,j)$-element of matrix $A$\\
 $A\succ0$ & $[A]_{i,j}>0$ $\forall i,j$ \\
$\mathcal M\subseteq \mathcal N$  & $\mathcal M$ is a subset of $\mathcal N$\\
$\mathcal M\subset \mathcal N$ & $\mathcal M$ is a proper subset of $\mathcal N$\\
 $|\mathcal M|$ &  cardinality of set $\mathcal M$ \\
 \bottomrule
\end{tabular}
\end{center}
\label{Notations}
\end{table}

\section{STP and Vector Representation of Logic}\label{SectionSTP}

\begin{definition}[Semi-tensor Product\cite{cheng2011analysis}]\label{DefinitonSTP}
   Let $A$ and $B$ be $m\times n$ and $p\times q$ matrices, respectively. The STP of $A$ and $B$ is defined as
 \[A\ltimes B :=(A\otimes I_{\alpha/n})(B\otimes I_{\alpha/p}),\]
where $ \alpha$ is the least common multiple of $n$ and $p$, $I_k$ denotes the $k \times k$ identity matrix, and ``$\otimes$'' represents the Kronecker product.
\end{definition}


\begin{definition}[Khatri--Rao Product]
   Let $A$ and $B$ be $m\times n$ and $p\times n$ matrices, respectively. The Khatri--Rao (K-R) product of $A$ and $B$, denoted by $A*B$, is defined as
   \[\col_j(A*B):=\col_j(A)\ltimes\col_j(B),\quad j\in[1:n],\]
where $\col_j(\cdot)$ represents the $j$th column of a matrix.
   \end{definition}

   The STP is compatible with the conventional matrix product and multiplication of matrices by numbers. Therefore, we can omit the symbol ``$\ltimes$'' without confusion. The STP inherits most of the properties of the conventional product of matrices, including associative and distributive laws.

One of the most successful applications of STP is the analysis and synthesis of logic systems by using a vector representation of logic. We use $\mathscr D_n$ to denote the $n$-valued logic domain, that is,
$\mathscr D_n:=[0:n-1]=\{0,1,\cdots, n-1\}$.
The vector form of any $\alpha\in\mathscr D_n$, denoted by $ \alpha$, is then defined as the $(n-\alpha)$th column of the identity matrix $I_n$, that is,
	$ \alpha:=\ddelta_n^{n-\alpha}$.
Thus, in vector form, the logic domain $\mathscr D_n$ becomes
$\Delta_n:=\{\ddelta_n^n,\ddelta_n^{n-1},\cdots, \ddelta_n^1\}$. For a $k$-dimensional logic vector
	$\alpha=(\alpha_1,\cdots,\alpha_k)^\top$ with $\alpha_j\in\mathscr D_{n_j}$ and $ j\in[1:k]$, the vector form of $\alpha$ is defined as
$ \alpha:= \alpha_1\ltimes  \alpha_2\ltimes\cdots\ltimes  \alpha_k$.
	We define the index mapping $\idx_n: \Delta_n\rightarrow[1:n]$ as $\idx_n(\ddelta_n^i)=i$, $i\in[1:n]$. The subscript $n$ of $\idx_n$ is omitted if dimension $n$ is available from the context.
For a subset $\mathcal M\subseteq\Delta_n$, $\idx(\mathcal M)$ represents the set of indices of the states in $\mathcal M$, that is, $\idx(\mathcal M):=\{j\in[1:n]\bigm| \ddelta_n^j\in\mathcal M\}$.
An $n\times m$ matrix $L$ is referred to as a logic matrix if $\col_j(L)\in\Delta_n$ for any $j\in[1:m]$.
 The set of all $n\times m$ logic matrices is denoted by $\mathscr L_{n\times m}$. For convenience, the logic matrix $L\in\mathscr L_{n\times m}$ with $\col_j(L)=\ddelta_n^{i_j}$, $j\in[1:m]$, and $i_j\in[1:n]$, is denoted by $\ddelta_n[i_1,i_2,\cdots, i_m]$.
	
%
%
%
%

	\begin{proposition}\cite{cheng2011analysis}\label{TheoremCheng} For any logical function
		\[f:\mathscr D_{n_1}\times\mathscr D_{n_2}\times\cdots\times\mathscr D_{n_k}\rightarrow\mathscr D_m,\] there exists a unique logical matrix $L_f\in\mathscr L_{m\times (n_1n_2\cdots n_k)}$, referred to as the structural matrix of $f$, such that
		\[  f(\alpha_1,\cdots,\alpha_k)=L_f \alpha_1\cdots  \alpha_k.\]

	\end{proposition}

\begin{remark}The structural matrix of a given logic function $f$ can be calculated using the properties of the STP. Further details are available in a prior work \cite{cheng2011analysis}.

\end{remark}

\section{Definitions and Problems Setting}\label{SectionDefinitions}

The switched LDS considered herein is of the form \begin{equation}\label{LDS1}
x(t+1)=f_{\sigma(t)}(x(t))
\end{equation}
where $x\in\Delta_n$ represents the state, $\sigma:\mathbb Z_{\geq0}\rightarrow[1:m]$ is the uncertain switching signal, and $f_j: \Delta_n\rightarrow\Delta_n$, $j\in[1:m]$,  are the logic functions. In algebraic form, an LDS (\ref{LDS1}) is expressed as
\begin{equation}\label{LDS-AlgebraicForm1}
   x(t+1)=L_{\sigma(t)} x(t),
\end{equation}
where $L_j\in\mathscr L_{n\times n}$ is the structural matrix of the logic function $f_j$ for any $j\in[1:m]$. We use $\mathcal S$ to denote the set of all switching signals, that is, \[\mathcal S:=\{\sigma\bigm|\sigma:\mathbb Z_{\geq0}\rightarrow[1:m] \}.\] Each switching signal $\sigma\in\mathcal S$ is called a switching pattern. Similarly, for any nonnegative integer $k$, we use $\mathcal S_k$ to denote the set of all switching patterns over $[0:k]$, that is,
\[\mathcal S_k:=\{\sigma\bigm|\sigma:[0:k]\rightarrow[1:m] \}.\]

\subsection{Robust Stabilities under Uncertain Switching}

\begin{definition}[Robust Set Stability]\label{Definition3}
   Given a subset $\mathcal M\subseteq\Delta_{n}$, an LDS (\ref{LDS-AlgebraicForm1}) is said to be robustly stable with respect to $\mathcal M$, or robustly $\mathcal M$-stable for short, if for any initial state $x_0\in\Delta_n$ and any switching signal $\sigma\in\mathcal S$, there exists a positive integer $T(x_0,\sigma)$ such that
   \[x(t;x_0,\sigma)\in\mathcal M\quad\forall t> T(x_0,\sigma).\]

   \end{definition}

   \begin{definition}[Uniform Robust Set Stability]\label{DefinitionURSS}
   Given a subset $\mathcal M\subseteq\Delta_n$, an LDS (\ref{LDS-AlgebraicForm1}) is said to be uniformly robustly stable with respect to $\mathcal M$ (or uniformly robustly $\mathcal M$-stable for short), if there is a positive integer $T$ such that for any initial state $ x_0\in\Delta_n$ and any switching signal $\sigma\in\mathcal S$,
   \[x(t;x_0,\sigma)\in\mathcal M\quad\forall t> T.\]
   \end{definition}

   \begin{remark} Robust set stability requires that each solution converges to the destination set within a finite time, whereas the uniform robust set stability additionally requires the transient periods to be uniformly finite with respect to arbitrary switching signals. In the next section, we prove that, although the uniform robust set stability implies robust set stability, the inverse implication is generally false.
   \end{remark}


    \begin{definition}[Reachability and Self-reachability]We consider LDS (\ref{LDS-AlgebraicForm1}).
   For any given integer $k\in\mathbb Z_{\geq1}$, a state $x_1\in\Delta_n$ of an LDS is said to be $k$-step reachable from $x_0\in\Delta_n$ if there is a switching signal $\sigma\in\mathcal S_{k-1}$  such that
   \[x(k;x_0,\sigma)=x_1.\]
In this case, we refer to $\sigma$ as a $k$-step reachable switching pattern from $ x_0$ to $x_1$.
   A state $x_1\in\Delta_n$ is said to be reachable from $x_0\in\Delta_n$ if it is $k$-step reachable from $x_0$  for some $k\in\mathbb Z_{\geq1}$. Otherwise, we state that $x_1$ is unreachable from $x_0$.
   A state is said to be self-reachable if it is reachable from itself.
   Otherwise, we state that it is self-unreachable.
\end{definition}

For convenience, we define
 \[ \boldsymbol\Gamma:=\frac{1}{m}(L_1+L_2+\cdots+L_m),
\quad\mathbf R:= \boldsymbol\Gamma+\boldsymbol\Gamma^2+\cdots +\boldsymbol\Gamma^n. \]
$\mathbf R$ is called the reachability matrix of the LDS (\ref{LDS-AlgebraicForm1}). It is easily checked that for any $i,j\in[1:n]$, $\ddelta_n^i$ is $k$-step reachable from $\ddelta_n^j$ if and only if $[\boldsymbol\Gamma^k]_{i,j}>0$. In addition, $\ddelta_n^i$ is reachable from $\ddelta_n^j$ if and only if $[\mathbf R]_{i,j}>0$.

The STG of LDS (\ref{LDS-AlgebraicForm1}) is a directed graph $(\mathcal V,\mathcal E)$ where $\mathcal V=\Delta_n$ is the set of vertices and $\mathcal E$ is the set of directed edges defined by
\[\mathcal E:=\{(x,y)\in\Delta_n\times\Delta_n\bigm| y^\top \boldsymbol\Gamma x>0\}.\]
A chain of states $x_0\rightarrow x_1\rightarrow\cdots\rightarrow x_k$, $k\in\mathbb Z_{\geq1}$, is called a path of the STG $(\mathcal V,\mathcal E)$ from $x_0$ to $x_k$ if $(x_{i-1},x_i)\in\mathcal E $ $\forall i\in[1:k]$. A path from a state to itself is termed a loop.
Evidently,  state $x_1$ is reachable from $x_0$ if and only if the STG has a path from $x_0$ to $x_1$. A state is self-reachable if and only if it is on a loop.

  \begin{definition}[Reachable Pattern Ratio]\label{DefinitionRotio}
     For any two states $x_0$, $x_1\in\Delta_n$ and any positive integer $k$, we denote by $\mathcal S_{k-1}(x_0,x_1)$ the set of all $k$-step reachable switching patterns from $x_0$ to $x_1$, that is,
     \[\mathcal S_{k-1}(x_0,x_1):=\left\{\sigma\in\mathcal S_{k-1}\bigm| x(k;x_0,\sigma)= x_1\right\}.\]
 Then, the $k$-step reachable pattern ratio from $x_0$ to $x_1$ is defined by
\[\boldsymbol\gamma_k(x_0,x_1):=\frac{|\mathcal S_{k-1}(x_0, x_1)|}{|\mathcal S_{k-1}|},\quad k\in\mathbb Z_{\geq1},\]
where $|\cdot|$ represents the cardinality of a set.
  \end{definition}

  For a subset $\mathcal M\subseteq\Delta_n$, we define
\begin{eqnarray*}
   \mathcal S_{k-1}(x_0,\mathcal M)&:=&\left\{\sigma\in\mathcal S_{k-1}\bigm| x(k;x_0,\sigma)\in \mathcal M\right\}=\bigcup_{x_1\in\mathcal M}\mathcal S_{k-1}(x_0,x_1).
\end{eqnarray*}
Then, the $k$-step reachable pattern ratio from $x_0$ to $\mathcal M$ is defined as
\[\boldsymbol\gamma_k(x_0,\mathcal M):=\frac{|\mathcal S_{k-1}(x_0,\mathcal M)|}{|\mathcal S_{k-1}|}=\sum_{x_1\in\mathcal M} \boldsymbol\gamma_k(x_0,x_1).\]

\begin{definition}[Asymptotical Stability with Ratio One]\label{DefinitionASRO}
   Given a subset $\mathcal M\subseteq\Delta_n$, an LDS (\ref{LDS-AlgebraicForm1}) is said to be asymptotically $\mathcal M$-stable with ratio one if
   \[\lim_{k\rightarrow\infty}\boldsymbol\gamma_k(x_0,\mathcal M)=1\quad\forall x_0\in\Delta_n.\]
\end{definition}

\begin{definition}[Finite-time Stability with Ratio One]\label{DefinitionFTSRO}
   Given a subset $\mathcal M\subseteq\Delta_n$, an LDS (\ref{LDS-AlgebraicForm1}) is said to be finite-time $\mathcal M$-stable with ratio one if there is a positive integer $K\in\mathbb Z_{\geq1}$ such that
   \[\boldsymbol\gamma_k(x_0,\mathcal M)=1\quad\forall x_0\in\Delta_n,\forall k\geq K.\]
\end{definition}

 \begin{remark}
By definition, finite-time stability with ratio one and uniformly robust stability are equivalent.
 \end{remark}

\subsection{Stabilities under Random Switching}

 In an LDS (\ref{LDS-AlgebraicForm1}),
if the switching signal $\sigma$ is an i.i.d. random sequence, the LDS(\ref{LDS-AlgebraicForm1}) becomes a  probabilistic logic system (PLS). If the probability distribution vector (PDV) of $\sigma(t)$ is $\boldsymbol p^\sigma$, which is a column vector satisfying
\[\mathbb P\{\sigma(t)=j\}=[\boldsymbol p^\sigma]_j,\quad j\in[1:m],\] the $1$-step transition probability of this PLS is
\begin{equation}
  \mathbf P(\boldsymbol p^\sigma):=\mathbf L\ltimes\boldsymbol p^\sigma,
\end{equation}
where
\[\mathbf L:=[L_1,L_2,\cdots, L_m].\]
As the stability of a PLS is completely determined by its TPM, for convenience, we simply use the TPM $\mathbf P(\boldsymbol p^\sigma)$ to denote this PLS. Without losing any generality, we assume that all components of the PDV $\boldsymbol p^\sigma$ are positive, that is, $\boldsymbol p^\sigma\succ0$.

\begin{definition}[Stability under Random Switching]
   LDS (\ref{LDS-AlgebraicForm1}) is said to be asymptotically $\mathcal M$-stable under random switching if PLS $\mathbf P(\boldsymbol p^\sigma)$ with $\boldsymbol p^\sigma\succ0$ is asymptotically $\mathcal M$-stable in distribution. It is said to be finite-time $\mathcal M$-stable under random switching if PLS $\mathbf P(\boldsymbol p^\sigma)$ with $\boldsymbol p^\sigma\succ0$ is finite-time $\mathcal M$-stable with probability one.
\end{definition}

\begin{remark}
   The problems of asymptotical and finite-time $\mathcal M$-stabilities of PLSs have been solved in previous studies \cite{Guo2019Stability,Li2014State}. Evidently, the asymptotical and finite-time $\mathcal M$-stabilities of PLS $\mathbf P(\boldsymbol p^\sigma)$ are independent of the PDV $\boldsymbol p^\sigma$ provided that $\boldsymbol p^\sigma\succ0$. Thus, the definition of stability under random witching is well-defined. We herein reveal the relations between the robust stability under uncertain switching and stability under random switching.
\end{remark}


\section{Robust Set Stability}\label{SectionRobustStability}

\begin{proposition}\label{Proposition1}
   LDS (\ref{LDS-AlgebraicForm1}) is robustly $\mathcal M$-stable if and only if all states in $\mathcal M^c:=\Delta_n\setminus\mathcal M$ are self-unreachable.
\end{proposition}

{\it Proof:}\quad (Sufficiency) Suppose that all states in $\mathcal M^c$ are self-unreachable. Taking any initial state $ x_0\in\Delta_n$ and any switching signal $\sigma\in\mathcal S$, based on the self-unreachability of all states in $\mathcal M^c$, any state in $\mathcal M^c$ can be visited once at most by the solution $ x(t; x_0,\sigma)$. We denote the set of all states that are visited by the solution $ x(t; x_0,\sigma)$ by $\mathcal V( x_0,\sigma)$, that is,
  \[\mathcal V(  x_0,\sigma):=\{ x_1\in\Delta_n\bigm| \exists t_1\in\mathbb Z_{\geq 0} \mbox{ s.t. }   x(t_1; x_0,\sigma)= x_1\},\]
and $T_{ x_1}( x_0,\sigma)$, the instant in time at which $ x_1\in \mathcal V( x_0,\sigma)$ is visited for the first time. For any $ x_0$ and $\sigma$, we define  $T( x_0,\sigma)$ as follows: \\
If $\mathcal M^c\cap\mathcal V( x_0,\sigma)\neq\emptyset$,
\[T( x_0,\sigma):=\max_{ x_1\in\mathcal M^c\cap\mathcal V( x_0,\sigma)}T_{ x_1}( x_0,\sigma).\]
Otherwise, $T( x_0,\sigma):=0$.
Then, it holds that $ x(t; x_0,\sigma)\in\mathcal M$ for any $t>T( x_0,\sigma)$. Thus, LDS (\ref{LDS-AlgebraicForm1}) is robustly $\mathcal M$-stable.

(Necessity) Supposing that LDS (\ref{LDS-AlgebraicForm1}) is robustly $\mathcal M$-stable and $ x_0$ is any state in $\mathcal M^c$, we show that $ x_0$ is self-unreachable. If $ x_0$ is self-reachable, there exist a switching signal $\sigma\in\mathcal S$ and a positive integer $k\geq1$ such that
$ x(k;\ x_0,\sigma)= x_0$.
We construct a periodic switching signal $\bar\sigma$ from $\sigma$ as
\[\bar\sigma(t):=\sigma(\mathbf r(t/k)),\quad\forall t\in\mathbb Z_{\geq0},\]
where $\mathbf r(t/k)$ denotes the remainder of $t$ divided by $k$.
Then, the solution $ x(t; x_0,\bar\sigma)$ is periodic and
\[ x(sk; x_0,\bar\sigma)= x_0\in\mathcal M^c\quad\forall s\in\mathbb Z_{\geq0}.\]
This contradicts the hypothesis of robust $\mathcal M$-stability.\hfill$\Box$

\begin{theorem}\label{Theorem1}
   LDS (\ref{LDS-AlgebraicForm1}) is robustly $\mathcal M$-stable if and only if
   \[[\mathbf R]_{i,i}=0,\quad\forall i\in\idx(\mathcal M^c).\]

\end{theorem}

{\it Proof:} This claim is a direct consequence of Proposition \ref{Proposition1} and the definition of $\mathbf R$.\hfill$\Box$

\begin{theorem}\label{Theorem2}
    LDS (\ref{LDS-AlgebraicForm1}) is robustly $\mathcal M$-stable if and only if all loops of the STG $(\mathcal V,\mathcal E)$ are in $\mathcal M$.
\end{theorem}

{\it Proof:} The claim directly follows Proposition \ref{Proposition1} and the fact that a state is self-reachable if and only if it is on a loop.\hfill $\Box$

\begin{corollary}\label{Corollary2}
   If $\mathcal M=\{\ddelta_n^j\}$ is a singleton, the following claims are equivalent:
    \begin{enumerate}[(a)]
       \item  LDS (\ref{LDS-AlgebraicForm1}) is robustly $\mathcal M$-stable;
       \item  $[\mathbf R]_{i,i}=0$ for any $i\in[1:n]$ with $i\neq j$; and
       \item  $\ddelta_n^j$ is a unique self-reachable state.
    \end{enumerate}

\end{corollary}

{\it Proof:} The equivalence of (a) and (b) is the direct consequence of Theorem \ref{Theorem1}. By Proposition \ref{Proposition1}, (c) implies (a). If (a) holds, $\ddelta_n^j$ is reachable from any state in $\Delta_n$, based on the definition of robust stability. Thus, $\ddelta_n^j$ is self-reachable. From Proposition \ref{Proposition1}, $\ddelta_n^j$ is a unique self-reachable state. \hfill$\Box$

\begin{remark}
     In \cite{2017Global}, a necessary and sufficient condition for the robust stability of a BN with respect to a fixed point was obtained. With slight modifications, the result obtained in \cite{2017Global} is also valid for general LDSs and is essentially equivalent to the conditions in Corollary \ref{Corollary2}. 
\end{remark}

\section{Uniform Robust Set Stability}\label{SectionUniformRobustStability}

\subsubsection{Reachability-based Results}

\begin{proposition}\label{Proposition3}
    LDS (\ref{LDS-AlgebraicForm1}) is uniformly robustly $\mathcal M$-stable if and only if all states in $\mathcal M^c$ are unreachable from any self-reachable state.
\end{proposition}

{\it Proof:\quad} (Sufficiency) Suppose that all states in $\mathcal M^c$ are unreachable from any self-reachable state. This hypothesis implies that all states in $\mathcal M^c$ are self-unreachable. Thus, by Proposition \ref{Proposition1}, LDS (\ref{LDS-AlgebraicForm1}) is robustly $\mathcal M$-stable. Next, we prove that it is also uniformly robustly $\mathcal M$-stable through a contradiction. If LDS (\ref{LDS-AlgebraicForm1}) is not uniformly robustly $\mathcal M$-stable, for any positive integer $T$, irrespective of its value, there exist an initial state $ x_{0,T}\in\Delta_n$, a switching signal $\sigma_T\in\mathcal S$, and an integer $k_T> T$, such that \[  x_{k,T}:= x(k_T; x_{0,T},\sigma_T)\in\mathcal M^c.\] We select a sufficiently large $T$ such that $T>n$, where $n$ denotes the cardinality of $\Delta_n$. Then, the solution $ x(t; x_{0,T},\sigma_T)$ must visit at least one self-reachable state within $[0:k_T-1]$.
 This implies that $ x_{k,T}$ is reachable from a self-reachable state, which contradicts the hypothesis.

(Necessity) Suppose that LDS (\ref{LDS-AlgebraicForm1}) is uniformly robustly $\mathcal M$-stable. Evidently, it is also robustly $\mathcal M$-stable. We prove that all states in $\mathcal M^c$ are unreachable from any self-reachable state through a contradiction. If there exists a self-reachable state $ x_0$ and a state $  x_1\in\mathcal M^c$ such that $  x_1$ is reachable from $ x_0$; in other words, there is a switching signal $\sigma$ and a positive integer $k_1$ such that $  x_1=x(k_1;  x_0,\sigma)$. Because $  x_0$ is self-reachable, there is a switching signal $\bar\sigma$ and a positive integer $\bar k$ such that $  x_0=x(\bar k;   x_0,\bar\sigma)$. We construct a sequence of switching signals $\sigma_j\in\mathcal S$, $j\in\mathbb Z_{\geq1}$ as
\[\sigma_j(t)=\left\{
    \begin{array}{ll}
         \bar\sigma(\mathbf r(t/\bar k)), & t/\bar k< j\\
         \sigma (t- j \bar k) &  t/\bar k \geq j.
    \end{array}
\right.\]
The following statement then holds:
\[  x(j\bar k+k_1;  x_0,\sigma_j)=  x_1\in\mathcal M^c\quad \forall j\in\mathbb Z_{\geq1}.\]
This implies that, for any given $T$ regardless of its size, there exist a $k>T$ and a switching signal $\sigma$ such that $  x(k;  x_0,\sigma)\in\mathcal M^c$; this contradicts the hypothesis that  LDS (\ref{LDS-AlgebraicForm1}) is uniformly robustly $\mathcal M$-stable. \hfill$\Box$

\begin{theorem}\label{Theorem3}

   LDS (\ref{LDS-AlgebraicForm1}) is uniformly robustly $\mathcal M$-stable if and only if
   \[[\mathbf R]_{i,j}=0,\quad \forall i\in\idx(\mathcal M^c),\;\forall j\in\idx(\mathcal C_0), \]
   where $\mathcal C_0$ is the set of self-reachable states, that is,
   \[\idx(\mathcal C_0)=\{s\in[1:n]\bigm|[\mathbf R]_{s,s}=1 \}.\]

\end{theorem}

{\it Proof:} The claim directly follows Proposition \ref{Proposition3} and the definition of $\mathbf R$.\hfill$\Box$

\begin{theorem}\label{Theorem4}

    LDS (\ref{LDS-AlgebraicForm1}) is uniformly robustly $\mathcal M$-stable if and only if every state on a loop has no path to $\mathcal M^c$.
\end{theorem}

{\it Proof:} This claim directly follows Proposition \ref{Proposition3} and the fact that the reachability from one state to another is equivalent to the existence of a path in STG $(\mathcal V,\mathcal E)$ connecting these two states.\hfill$\Box$

\begin{corollary}\label{Corollary1}
   If $\mathcal M$ is a singleton, LDS (\ref{LDS-AlgebraicForm1}) is uniformly robustly $\mathcal M$-stable if and only if it is robustly $\mathcal M$-stable.

\end{corollary}

{\it Proof:} We only need to prove the sufficiency. The state space $\Delta_n$ is finite; thus, LDS (\ref{LDS-AlgebraicForm1}) has at least one self-reachable state. If LDS (\ref{LDS-AlgebraicForm1}) is robustly stable with respect to $\mathcal M=\{  x_0\}$, by Corollary \ref{Corollary2}, $  x_0$ is a unique self-reachable state. Then, for any $  x_1\neq  x_0$, there is no path from $  x_0$ to $  x_1$; otherwise, $  x_1$ is also self-reachable. By Theorem \ref{Theorem4}, LDS (\ref{LDS-AlgebraicForm1}) is uniformly robustly stable with respect to $\{  x_0\}$.\hfill$\Box$

\subsubsection{LRIS-based Results}

\begin{definition}({\bf Robustly Invariant Subset})
  A subset $\mathcal{C}\subseteq\Delta_n$ is said to be a robustly invariant subset of an LDS (\ref{LDS-AlgebraicForm1}) if
      \[ x(t; x_0,\sigma)\in\mathcal{C}\quad \forall  x_0\in\mathcal{C}, \forall \sigma\in\mathcal S, \forall t\geq0.\]
 \end{definition}

For any given subset $\mathcal M\subseteq \Delta_n$, the union of any two robustly invariant subsets in $\mathcal{M}$ is still robustly invariant. We denote by $I(\mathcal{M})$ the union of all robustly invariant subsets in $\mathcal{M}$, which is referred to as the LRIS in $\mathcal M$.

\begin{lemma}\label{lemma1}
    If LDS (\ref{LDS-AlgebraicForm1}) is uniformly robustly $\mathcal M$-stable, $I(\mathcal M)$ contains all loops of STG $(\mathcal V,\mathcal E)$.
\end{lemma}

{\it Proof:} By Theorem \ref{Theorem2} and the hypothesis of uniform robust stability, $\mathcal M$ contains all loops. According to Theorem \ref{Theorem4}, all states on loops have no paths to $\mathcal M^c$. This implies that any solution starting from a state on a loop never escapes from $\mathcal M$ under arbitrary switching. According to the definition of the LRIS, all states on loops are in $I(\mathcal M)$.\hfill$\Box$

\begin{lemma}\label{lemma2}
   Suppose that $\mathcal C$ is a robustly invariant subset. Then, LDS (\ref{LDS-AlgebraicForm1}) is uniformly robustly $\mathcal C$-stable if and only if it is robustly $\mathcal C$-stable.
\end{lemma}

{\it Proof:} This necessity is clearly true. We prove the sufficiency in the following. If LDS (\ref{LDS-AlgebraicForm1}) is robustly $\mathcal C$-stable, by Theorem \ref{Theorem2}, all loops of STG $(\mathcal V,\mathcal E)$ are in $\mathcal C$. Based on this observation and the definition of a robustly invariant subset, any state that is on a loop has no path to $\mathcal C^c$. By Theorem \ref{Theorem4}, LDS (\ref{LDS-AlgebraicForm1}) is uniformly robustly $\mathcal C$-stable.\hfill$\Box$

\begin{theorem}\label{Theorem5}
The following claims are equivalent:
\begin{enumerate}[(a)]
  \item LDS (\ref{LDS-AlgebraicForm1}) is uniformly robustly $\mathcal M$-stable;
   \item LDS (\ref{LDS-AlgebraicForm1}) is uniformly robustly $I(\mathcal M)$-stable; and
      \item LDS (\ref{LDS-AlgebraicForm1}) is robustly $I(\mathcal M)$-stable.
\end{enumerate}

\end{theorem}

{\it Proof:} By Lemma \ref{lemma2}, claims (b) and (c) are equivalent. In addition, (b) implies (a) because $I(\mathcal M)\subseteq\mathcal M$. Next, we prove that (a) implies (b). Suppose that LDS (\ref{LDS-AlgebraicForm1}) is uniformly robustly $\mathcal M$-stable. Then, there exists a positive integer $T$ such that for any initial state $  x_0\in\Delta_n$ and any switching signal $\sigma\in\mathcal S$, the following holds: $  x(t;  x_0,\sigma)\in\mathcal M$ for any $t>T$. Without any loss of generality, we can assume that $T>n$. Thus, the segment of any solution $  x(t;  x_0,\sigma)$, $t\in[0:T]$, contains at least one loop. From Lemma \ref{lemma1},
 any solution must enter $I(\mathcal M)$ within $T$ steps. Based on the definition of the LRIS, the following holds:
\[  x(t;  x_0,\sigma)\in I(\mathcal M)\quad \forall t>T, \forall   x_0\in\Delta_n, \forall \sigma\in\mathcal S.\]
This implies that LDS (\ref{LDS-AlgebraicForm1}) is uniformly robustly $I(\mathcal M)$-stable.\hfill $\Box$

\begin{remark}

    Notably, a uniform stability is crucial for establishing the LRIS-based criterion in Theorem \ref{Theorem5}. If an LDS is robustly $ \mathcal M$-stable, it is not necessarily robustly $I(\mathcal M)$-stable, as shown in Example \ref{Example1} in Section \ref{SubSecRelations}.
    The robust set stability proposed herein is based on the definition of set stability proposed in \cite{Guo2017Invariant}. Thus, the necessity of Theorem 1 in \cite{Guo2017Invariant} is incorrect and should be corrected by replacing the definition of set stability with the uniform robust set stability proposed in this study.
\end{remark}

\begin{remark}\label{remark1}
If an LDS suffers no uncertainties, that is, there is only one subnetwork, the robust set stability then degenerates to the ordinary set stability proposed in \cite{Guo2015Set}. In other words, the solution starting from any initial state converges to the destination set in finite time. In this case, observing that the total number of initial states is finite, we can always find a time step that is uniformly finite with respect to all initial states, within which any solution converges to the destination set. This implies that the uniform robust set stability also degenerates to the ordinary set stability in the uncertainty-free case. This observation reveals that the uncertainty causes an essential difference between the robust set stability and uniform robust set stability.
\end{remark}

 \section{Stability with Ratio One}\label{SectionStabilityRatioOne}

 \subsection{Reachable Pattern Ratio Matrix}

 The $k$-step reachable pattern ratio matrix (TRM) is a $n\times n$ matrix $\boldsymbol\Gamma(k)$ defined as
  \[[\boldsymbol\Gamma(k)]_{i,j}:=\boldsymbol\gamma_k(\ddelta_n^j,\ddelta_n^i),\quad i,j\in[1:n].\]
  The following lemma establishes the equivalence between the stability of uncertain LDSs with ratio one and the stability of PLSs with probability one.

\begin{lemma}\label{Lemma220301-1}
For any positive integer $k\in\mathbb Z_{\geq1}$, the $k$-step reachable pattern ratio matrix $\boldsymbol\Gamma^\top(k)$ is a stochastic matrix, that is, \[0\leq[\boldsymbol\Gamma(k)]_{i,j}\leq1\quad\forall i,j\in[1:N],\]
 \[\sum_{i\in[1:N]}[\boldsymbol\Gamma(k)]_{i,j}=1.\]
 In addition, the following holds:
 \begin{equation}\label{Eqn220301-2}
    \boldsymbol\Gamma(k)=\boldsymbol\Gamma^k,\quad\forall k\in\mathbb Z_{\geq1}.
 \end{equation}

\end{lemma}

{\it Proof:} The claim that $\boldsymbol\Gamma^\top(k)$ is a stochastic matrix for any positive integer $k$ is evidently true by definition. Notably, $|\mathcal S_{k-1} |=m^k$ $\forall k\in\mathbb Z_{\geq 1}$. Thus, by the definitions of $\boldsymbol\gamma_k(x_0,x_1)$ and $\boldsymbol\Gamma$, in order to prove (\ref{Eqn220301-2}), we only need to prove
\begin{equation}\label{Eqn220301-1}
   |\mathcal S_{k-1}(\ddelta_n^j,\ddelta_n^i)|=[\mathbf Q^k]_{i,j}\quad\forall k\in\mathbb Z_{\geq1},
\end{equation}
where \[\mathbf Q:=L_1+L_2+\cdots+L_M=m\boldsymbol\Gamma.\]
We can prove (\ref{Eqn220301-1}) by induction.
 It is verifiable that (\ref{Eqn220301-1}) holds for $k=1$, that is,
\[|\mathcal S_0(\ddelta_n^j,\ddelta_n^i)|=[\mathbf Q]_{i,j}.\]
Suppose that (\ref{Eqn220301-1}) holds for $k=s$, that is,
\[|\mathcal S_{s-1}(\ddelta_n^j,\ddelta_n^i)|=[\mathbf Q^s]_{i,j}\quad\forall i,j\in[1:n].\]
Then, for $k=s+1$,
\begin{eqnarray*}
  |\mathcal S_{s}(\ddelta_n^j,\ddelta_n^i)|
   &=& \sum_{r\in[1:n]}|\mathcal S_{s-1}(\ddelta_n^j,\ddelta_n^r)|   |\mathcal S_0(\ddelta_n^r,\ddelta_n^i)|\\
   &=&\sum_{r\in[1:n]} [\mathbf Q^s]_{r,j}[\mathbf Q]_{i,r}\\
   &=&[\mathbf Q^{s+1}]_{i,j}.
\end{eqnarray*}
\hfill $\Box$

\begin{proposition}For a given subset $\mathcal M\subseteq\Delta_n$, we denote by $\boldsymbol\beta_{\mathcal M}$ the Boolean vector, which is defined as
\[\boldsymbol\beta_{\mathcal M}:=\sum_{x\in\mathcal M}x.\]
Then, the following claims hold:
\begin{enumerate}
  \item  LDS (\ref{LDS-AlgebraicForm1}) is asymptotically $\mathcal M$-stable with ratio one if and only if
      \[\lim_{k\rightarrow\infty} \boldsymbol\beta_{\mathcal M}^\top\boldsymbol\Gamma^k=\mathbf 1_n^\top.\]
   \item  LDS (\ref{LDS-AlgebraicForm1}) is finite-time $\mathcal M$-stable with ratio one if and only if there is a positive integer $K$ such that
             \[  \boldsymbol\beta_{\mathcal M}^\top\boldsymbol\Gamma^k=\mathbf 1_n^\top\quad\forall k\geq K.\]
\end{enumerate}
\end{proposition}

{\it Proof:}  According to Lemma \ref{Lemma220301-1} and the definition of $\boldsymbol\gamma_k(x_0,\mathcal M)$, the following holds:
\[\boldsymbol\gamma_k(x_0,\mathcal M)=\sum_{x_1\in\mathcal M} \boldsymbol\gamma_k(x_0,x_1)=\boldsymbol\beta_{\mathcal M}^\top\boldsymbol\Gamma(k)x_0=\boldsymbol\beta_{\mathcal M}^\top\boldsymbol\Gamma^kx_0.\]
On this basis and given the Definitions \ref{DefinitionASRO} and \ref{DefinitionFTSRO}, the claims follow directly.\hfill$\Box$

\subsection{Asymptotical Stability with Ratio One}

\begin{proposition}\label{Prop220320-1}
 LDS (\ref{LDS-AlgebraicForm1}) is asymptotically $\mathcal M$-stable with ratio one if and only if it is asymptotically $\mathcal M$-stable under random switching.

\end{proposition}

{\it Proof:} The asymptotical stability of  LDS (\ref{LDS-AlgebraicForm1}) under random switching is independent of the PDV $\boldsymbol p^\sigma$ provided that $\boldsymbol p^\sigma\succ0$. Without losing any generality, we assume that \[\boldsymbol p^\sigma=\frac{1}{m}\mathbf 1_m.\]
In this case,
\[\mathbf P(\boldsymbol p^\sigma)= \frac{1}{m} \mathbf L\ltimes \mathbf 1_m =\boldsymbol\Gamma.\]
This implies that the $k$-step reachable pattern ratio of uncertain LDS (\ref{LDS-AlgebraicForm1}) coincides with the state transition probability of the PLS $\mathbf P(\boldsymbol p^\sigma)$, that is,
\[\mathbf P^k(\boldsymbol p^\sigma)=\boldsymbol \Gamma^k,\quad k\in\mathbb Z_{\geq1}.\]
 By Lemma \ref{Lemma220301-1}, the reachable pattern ratio $\boldsymbol\gamma_k(x_0,\mathcal M)$ of LDS (\ref{LDS-AlgebraicForm1}) equals the $k$-step transition probability of PLS $\mathbf P(\boldsymbol p^\sigma)$ from $x_0$ to $\mathcal M$. By the definitions of asymptotical stability of PLSs, the claim follows.\hfill $\Box$

\begin{lemma}\label{Lemma220320-1}
Suppose that $\boldsymbol p^\sigma$ is any PDV satisfying
$\boldsymbol p^\sigma\succ0$. Then, a subset $\mathcal C\subseteq\Delta_n$ is a robustly invariant subset of LDS (\ref{LDS-AlgebraicForm1}) if and only if it is an invariant subset of PLS $\mathbf P(\boldsymbol p^\sigma)$.
\end{lemma}

{\it Proof:}  By the definition, a subset $\mathcal{C}\subseteq\Delta_n$ is a robustly invariant subset of an LDS (\ref{LDS-AlgebraicForm1}) if and only if
   \begin{equation}\label{Eqn220319-1}
      \boldsymbol\gamma_k(x_0,\mathcal C)=1\quad\forall x_0\in\mathcal C,\;\forall k\in\mathbb Z_{\geq1}.
   \end{equation}
   According to Lemma \ref{Lemma220301-1}, (\ref{Eqn220319-1}) is equivalent to
   \begin{equation}\label{Eqn220319-2}
      \sum_{i\in\idx(\mathcal C)}[\boldsymbol\Gamma]_{i,j}=1\quad\forall j\in\idx(\mathcal C).
   \end{equation}
   Notably, both $\boldsymbol\Gamma$ and $\mathbf P(\boldsymbol p^\sigma)$ are TPMs, and  the assumption $\boldsymbol p^\sigma\succ0$ implies that the zero element locations in $\boldsymbol\Gamma$ and $\mathbf P(\boldsymbol p^\sigma)$ are equal. Then, (\ref{Eqn220319-2}) holds if and only if
   \begin{equation}\label{Eqn220319-3}
      \sum_{i\in\idx(\mathcal C)}[\mathbf P(\boldsymbol p^\sigma)]_{i,j}=1\quad\forall j\in\idx(\mathcal C).
   \end{equation}
   This is equivalent to stating that $\mathcal C$ is an invariant subset of PLS $\mathbf P(\boldsymbol p^\sigma)$.\hfill $\Box$

\begin{remark}
   By Lemma \ref{Lemma220320-1}, for any given set $\mathcal M$, the LRIS $I(\mathcal M)$ is also the LIS of PLS $\mathbf P(\boldsymbol p^\sigma)$ in $\mathcal M$ provided that $\boldsymbol p^\sigma\succ0$. Thus, for any given subset $\mathcal M$, $I(\mathcal M)$ can be calculated by using the algorithm proposed in \cite{Guo2019Stability}.
\end{remark}

   \begin{theorem}\label{theorem6}
   The following statements are equivalent:
   \begin{enumerate}
      \item LDS (\ref{LDS-AlgebraicForm1}) is asymptotically $\mathcal M$-stable with ratio one.
      \item LDS (\ref{LDS-AlgebraicForm1}) is asymptotically $I(\mathcal M)$-stable with ratio one.
      \item Every vertex of STG $(\mathcal V,\mathcal E)$ has a path to $I(\mathcal M)$, that is, $I(\mathcal M)\neq\emptyset$ and
       \begin{equation}\label{Eqn220320-1}
          \sum_{i\in\idx(I(\mathcal M))}[\mathbf R]_{i,j}>0\quad\forall j\in[1:n].
       \end{equation}
   \end{enumerate}

   \end{theorem}

   {\it Proof:}  By Proposition \ref{Prop220320-1}, the asymptotical $\mathcal M$-stability with ratio one  of LDS (\ref{LDS-AlgebraicForm1}) is equivalent to the asymptotical $\mathcal M$-stability in distribution of PLS $\mathbf P(\boldsymbol p^\sigma)$. The asymptotical $\mathcal M$-stability  is, in turn, equivalent to the asymptotical stability in distribution with respect to the LIS of PLS $\mathbf P(\boldsymbol p^\sigma)$ in $\mathcal M$, which is identical to the LRIS $I(\mathcal M)$ of LDS (\ref{LDS-AlgebraicForm1}) in $\mathcal M$ by Lemma \ref{Lemma220320-1}. By Theorem 3 in \cite{Guo2019Stability}, the equivalence between the statements follow.\hfill$\Box$

\subsection{Finite-time Stability with Ratio One}

   \begin{proposition}\label{Prop220322-1}
Suppose that $\boldsymbol p^\sigma$ is any PDV of $\sigma(t)$ satisfying $\boldsymbol p^\sigma\succ0$. Then, the following statements are equivalent:
       \begin{enumerate}
         \item  LDS (\ref{LDS-AlgebraicForm1}) is finite-time $\mathcal M$-stable with ratio one.
         \item LDS (\ref{LDS-AlgebraicForm1}) is uniformly robustly  $\mathcal M$-stable
         \item LDS (\ref{LDS-AlgebraicForm1}) is finite-time $\mathcal M$-stable under random switching.
       \end{enumerate}
   \end{proposition}

   {\it Proof:}  By the definitions, the finite-time stability with ratio one and uniformly robust stability are equivalent. Thus, according to Proposition \ref{Proposition3}, statements 1) and 2) are equivalent. As regards the equivalence between statements 1) and 3), we first note that the finite-time stability with probability one of PLS $\mathbf P(\boldsymbol p^\sigma)$ is completely determined by the locations of nonzero elements in $\mathbf P(\boldsymbol p^\sigma)$, which are the same as those in $\boldsymbol\Gamma$ provided that $\boldsymbol p^\sigma\succ0$. Then, by following the same argument in the proof of Proposition \ref{Prop220320-1}, the claim follows.\hfill$\Box$

   \begin{theorem}\label{Theorem220322-1}
       The following statements are equivalent:
       \begin{enumerate}
         \item LDS (\ref{LDS-AlgebraicForm1}) is finite-time $\mathcal M$-stable with ratio one.
         \item All states in $\mathcal M^c$ are unreachable from any self-reachable state, that is,
             \[[\mathbf R]_{i,j}=0,\quad \forall i\in\idx(\mathcal M^c),\;\forall j\in\idx(\mathcal C_0), \]
   where $\mathcal C_0$ is the set of self-reachable states.
         \item The STG $(\mathcal V,\mathcal E)$ has no loops outside $I(\mathcal M)$, or equivalently, all states outside $I(\mathcal M)$ is self-unreachable, that is,
                \[[\mathbf R]_{i,i}=0,\quad\forall i\in[1:n]\setminus\idx(I(\mathcal M)).\]
       \end{enumerate}
   \end{theorem}

   {\it Proof:} The equivalence between 1) and 2) is a direct consequence of Proposition \ref{Prop220322-1} and Theorem \ref{Theorem3}. By Theorem \ref{Theorem5} and Proposition \ref{Prop220322-1}, LDS (\ref{LDS-AlgebraicForm1}) is finite-time $\mathcal M$-stable with ratio one if and only if it is robustly $I(\mathcal M)$-stable. Thus, by Theorem \ref{Theorem1}, 1) and 3) are equivalent.\hfill $\Box$

   By using Theorem \ref{Theorem220322-1} and Proposition \ref{Prop220322-1}, we have the following corollary.

   \begin{corollary}\label{corollary220322-1}
       PLS $\mathbf P(\boldsymbol p^\sigma)$ is finite-time $\mathcal M$-stable with probability one if and only if its STG has  no loops outside $I(\mathcal M)$.
   \end{corollary}

   \begin{remark}
      It has been proven in \cite{zhu2020finite} that a PLS is finite-time stable with probability one with respect to a fixed point if and only if its STG has no additional loops expect for the self-loop of the fixed point. Corollary \ref{corollary220322-1} essentially generalized this result to the finite-time stability of PLSs with respect to subsets.
   \end{remark}

\section{Relations between Different Robust Stabilities}\label{SubSecRelations}

%

 Theorem \ref{Theorem5} shows that the uniform robust set stability, uniform robust stability with respect to the LRIS, and robust stability with respect to the LRIS are equivalent. In addition, if an LDS (\ref{LDS-AlgebraicForm1}) is uniformly robustly $ \mathcal M$-stable, it is also robustly $ \mathcal M$-stable according to the definitions; however, the reverse claim is not generally true, as shown by  Example \ref{Example1} in Section \ref{SubSectionExample}. However, we proved that the robust set stability and uniform robust set stability become equivalent under one of the following three situations:
 \begin{enumerate}[(a)]
    \item The destination set $\mathcal M$ is a singleton (Corollary \ref{Corollary1}).
    \item The destination set $\mathcal M$ is robustly invariant (Lemma \ref{lemma2}).
    \item The LDS suffers from no uncertainty, that is, there is only one subnetwork (Remark \ref{remark1}).
 \end{enumerate}

 Proposition \ref{Prop220322-1} shows that the uniform robust set stability and finite-time set stability with ratio one are equivalent. The following proposition affords the relation between robust set stability and asymptotical set stability with ratio one.

 \begin{proposition}\label{Proposition7}
     If an LDS (\ref{LDS-AlgebraicForm1}) is robustly $\mathcal M$-stable, it is also asymptotically $\mathcal M$-stable with ratio one.
 \end{proposition}

 {\it Proof:} If LDS (\ref{LDS-AlgebraicForm1}) is robustly $\mathcal M$-stable, by Definition \ref{Definition3}, for any initial state $x_0\in\Delta_n$ and any switching signal $\sigma\in\mathcal S$, there exists a positive integer $T(x_0,\sigma)$ such that
   \[x(t;x_0,\sigma)\in\mathcal M\quad\forall t> T(x_0,\sigma).\]
  This implies that for any $t_1 >T(x_0,\sigma)$,
  \[x_1:=x(t_1;x_0,\sigma)\in I(\mathcal M).\]
  Therefore, $I(\mathcal M)\neq\emptyset$ and any state has a path to $I(\mathcal M)$. By Theorem \ref{theorem6}, LDS  (\ref{LDS-AlgebraicForm1}) is asymptotically $\mathcal M$-stable with ratio one.\hfill$\Box$

 \begin{remark}
     The reverse claim of Proposition \ref{Proposition7} does not hold, that is, asymptotical set stability with ratio one does not imply robust set stability, as shown in Example \ref{Example3} in Section \ref{SubSectionExample}.
 \end{remark}

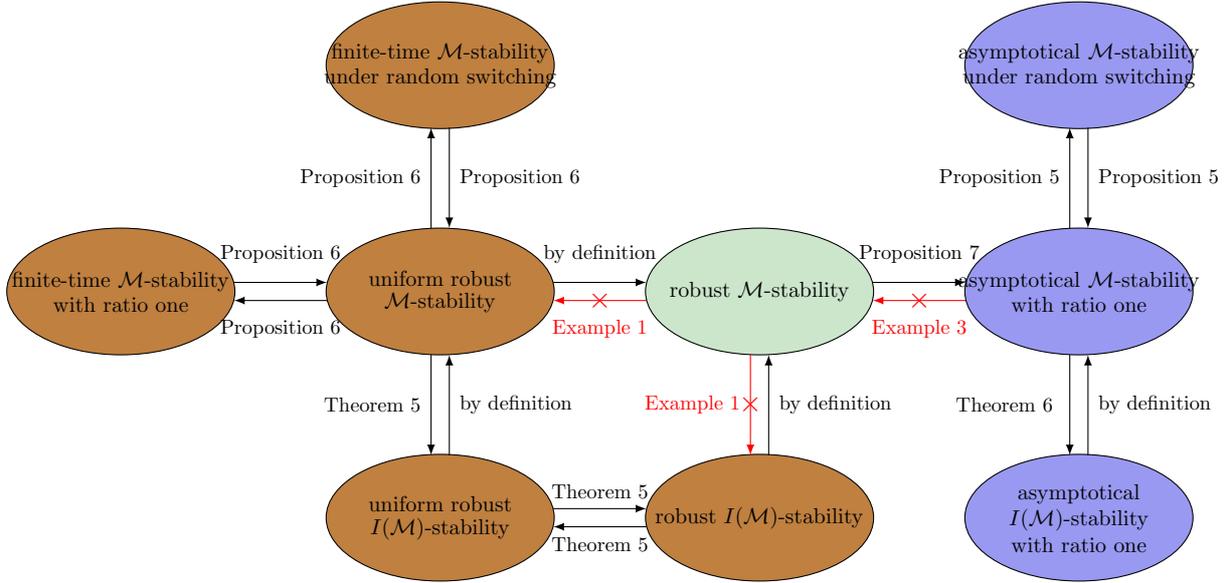
\begin{figure*}
\begin{center}
   \begin{tikzpicture}[>=latex, scale=1.2]
\draw[fill=brown] (0,0) ellipse (1.25 and 0.7);
\coordinate [label=center:{\scalebox{0.75}{\shortstack{finite-time $\mathcal M$-stability \\ with ratio one }}}] (B) at (0,0);

\draw[->,-latex](1.24,0.1)--(2.26,0.1);
\coordinate [label=above:{\scalebox{0.7}{Proposition \ref{Prop220322-1}}}] (B) at (1.75,0.2);

\draw[->,-latex](2.26,-0.1)--(1.24,-0.1);
\coordinate [label=below:{\scalebox{0.7}{Proposition \ref{Prop220322-1}}}] (B) at (1.75,-0.2);

\draw[fill=brown] (3.5,0) ellipse (1.25 and 0.7);
\coordinate [label=center:{\scalebox{0.75}{\shortstack{uniform robust\\$\mathcal M$-stability}}}] (B) at (3.5,0);

\draw[->,-latex](4.74,0.1)--(5.76,0.1);
\coordinate [label=above:{\scalebox{0.7}{by definition}}] (B) at (5.25,0.2);

\draw[->,-latex,color=red](5.76,-0.1)--(4.74,-0.1);
\coordinate [label=center:{\scalebox{1}{\color{red}$\times$}}] (B) at (5.25,-0.1);

\coordinate [label=below:{\scalebox{0.7}{\color{red}Example \ref{Example1}}}] (B) at (5.25,-0.2);

\draw[fill=LGreen] (7,0) ellipse (1.25 and 0.7);
\coordinate [label=center:{\scalebox{0.75}{\shortstack{robust $\mathcal M$-stability }}}] (B) at (7,0);

\draw[->,-latex](8.24,0.1)--(9.26,0.1);
\coordinate [label=above:{\scalebox{0.7}{Proposition \ref{Proposition7}}}] (B) at (8.75,0.2);

\draw[->,-latex,color=red](9.26,-0.1)--(8.24,-0.1);
\coordinate [label=center:{\scalebox{1}{\color{red}$\times$}}] (B) at (8.75,-0.1);

\coordinate [label=below:{\scalebox{0.7}{\color{red}Example \ref{Example3}}}] (B) at (8.75,-0.2);

\draw[fill=LBlue] (10.5,0) ellipse (1.25 and 0.7);
\coordinate [label=center:{\scalebox{0.75}{\shortstack{asymptotical $\mathcal M$-stability \\ with ratio one }}}] (B) at (10.5,0);

\draw[fill=brown] (3.5,2.5) ellipse (1.25 and 0.7);
\coordinate [label=center:{\scalebox{0.75}{\shortstack{finite-time $\mathcal M$-stability\\under random switching}}}] (B) at (3.5,2.5);

\draw[->,-latex](3.4,0.7)--(3.4,1.81);
\coordinate [label=left:{\scalebox{0.7}{Proposition \ref{Prop220322-1}}}] (B) at (3.4,1.25);

\draw[->,-latex](3.6,1.81)--(3.6,0.7);
\coordinate [label=right:{\scalebox{0.7}{Proposition \ref{Prop220322-1}}}] (B) at (3.6,1.25);

\draw[fill=LBlue]  (10.5,2.5) ellipse (1.25 and 0.7);
\coordinate [label=center:{\scalebox{0.75}{\shortstack{asymptotical $\mathcal M$-stability\\under random switching}}}] (B) at (10.5,2.5);

\draw[->,-latex](10.4,0.7)--(10.4,1.81);
\coordinate [label=left:{\scalebox{0.7}{Proposition \ref{Prop220320-1}}}] (B) at (10.4,1.25);

\draw[->,-latex](10.6,1.81)--(10.6,0.7);
\coordinate [label=right:{\scalebox{0.7}{Proposition \ref{Prop220320-1}}}] (B) at (10.6,1.25);

\draw[fill=brown]  (3.5,-2.5) ellipse (1.25 and 0.7);
\coordinate [label=center:{\scalebox{0.75}{\shortstack{uniform robust\\$I(\mathcal M)$-stability }}}] (B) at (3.5,-2.5);

\draw[->,-latex](3.4,-0.7)--(3.4,-1.81);
\coordinate [label=left:{\scalebox{0.7}{Theorem \ref{Theorem5}}}] (B) at (3.4,-1.25);

\draw[->,-latex](3.6,-1.81)--(3.6,-0.7);
\coordinate [label=right:{\scalebox{0.7}{by definition}}] (B) at (3.6,-1.25);

\draw[fill=brown]  (7,-2.5) ellipse (1.25 and 0.7);
\coordinate [label=center:{\scalebox{0.75}{\shortstack{ robust $I(\mathcal M)$-stability }}}] (B) at (7,-2.5);

\draw[->,-latex,color=red](6.9,-0.7)--(6.9,-1.81);
\coordinate [label=center:{\scalebox{1}{\color{red}$\times$}}] (B) at (6.9,-1.25);
\coordinate [label=left:{\scalebox{0.7}{\color{red}Example \ref{Example1}}}] (B) at (6.9,-1.25);

\draw[->,-latex](7.1,-1.81)--(7.1,-0.7);
\coordinate [label=right:{\scalebox{0.7}{by definition}}] (B) at (7.1,-1.25);

\draw[->,-latex](4.74,-2.4)--(5.76,-2.4);
\coordinate [label=above:{\scalebox{0.7}{Theorem \ref{Theorem5}}}] (B) at (5.25,-2.4);

\draw[->,-latex](5.76,-2.6)--(4.74,-2.6);
 \coordinate [label=below:{\scalebox{0.7}{Theorem \ref{Theorem5}}}] (B) at (5.25,-2.6);

\draw[fill=LBlue]  (10.5,-2.5) ellipse (1.25 and 0.7);
\coordinate [label=center:{\scalebox{0.75}{\shortstack{asymptotical \\$I(\mathcal M)$-stability\\with ratio one}}}] (B) at (10.5,-2.5);

\draw[->,-latex](10.4,-0.7)--(10.4,-1.81);
\coordinate [label=left:{\scalebox{0.7}{Theorem \ref{theorem6} }}] (B) at (10.4,-1.25);

\draw[->,-latex](10.6,-1.81)--(10.6,-0.7);
\coordinate [label=right:{\scalebox{0.7}{by definition }}] (B) at (10.6,-1.25);

\end{tikzpicture}
\end{center}
    \caption{Relations between different stabilities. Each arrow represents an ``implication'' and a crossed arrow represents ``a non-implication''; the  ellipses in the same color represent equivalent definitions of stability}
\label{Figure2}
\end{figure*}

 The relations between these robust stabilities are shown in Figure \ref{Figure2}.

\subsection{Illustrative Examples}\label{SubSectionExample}

\begin{example}\label{Example1}
    Consider an LDS (\ref{LDS-AlgebraicForm1}) with $N = 8$, $M = 2$, and
    \[L_1=\ddelta_8[2,3,4,3,8,7,6,7],\quad L_2=\ddelta_8[5,5,5,3,6,7,7,6].\]
     We can determine the robust stability and uniform robust stability with respect to $\mathcal M$ with
          \[\mathcal M=\{\ddelta_8^3,\ddelta_8^4,\ddelta_8^6,\ddelta_8^7,\ddelta_8^8\}.\]
We calculate the reachability matrix as follows:
\begin{eqnarray*}
   \boldsymbol\Gamma&=&\frac{1}{2}(L_1+ L_2)=\left[
    \begin{array}{cccccccc}
       0&0&0&0&0&0&0&0\\
       0.5&0&0&0&0&0&0&0\\
       0&0.5&0&1&0&0&0&0\\
       0&0&0.5&0&0&0&0&0\\
       0.5&0.5&0.5&0&0&0&0&0\\
       0&0&0&0&0.5&0&0.5&0.5\\
       0&0&0&0&0&1&0.5&0.5\\
       0&0&0&0&0.5&0&0&0
    \end{array}
\right],
\end{eqnarray*}
\begin{eqnarray*}
   \mathbf R&=&\boldsymbol\Gamma+ \boldsymbol\Gamma^2+ \cdots+\boldsymbol\Gamma^8\approx\left[
     \begin{array}{cccccccc}
         0    &     0   &      0   &      0  &        0  &        0    &      0 &         0\\
    0.50 &      0    &      0   &       0  &        0    &      0   &       0    &      0\\
    0.47  &   0.94 &     0.94 &    1.88    &      0    &      0   &       0    &      0\\
    0.22 &    0.47  &   0.94  &   0.94     &     0  &        0    &      0   &       0\\
    0.97 &    0.97  &   0.94  &   0.94     &     0   &       0    &      0  &        0\\
    2.06  &   1.98  &   1.84  &   1.51  &   2.78  &   2.45  &   2.78  & 2.78\\
    3.30 &   3.18 &    2.88 &   2.30 &     4.72 &   5.55 &     5.22 &     5.22\\
    0.48 &    0.47 &   0.47 &   0.44 &    0.50&          0  &        0   &       0
     \end{array}
\right].
\end{eqnarray*}
It is verifiable that
\[[\mathbf R]_{i,i}=0\quad\forall i\in\idx(\mathcal M^c)=\{1,2,5\}.\]
Thus, by Theorem \ref{Theorem1}, this LDS is robustly $\mathcal M$-stable. In addition, $[\mathbf R]_{5,j}>0$ for $j = 3,4$. However,  the set of self-reachable states is
\begin{eqnarray*}
   \mathcal C_0&:=&\{\ddelta_8^j\bigm|[\mathbf R]_{j,j} >0\}=\{\ddelta_8^j\bigm| j=3,4,6,7\}.
\end{eqnarray*}
Thus, the state $\ddelta_8^5$, which is in $\mathcal M^c$, is reachable from the self-reachable states $\ddelta_8^3$ and $\ddelta_8^4$. By Theorem \ref{Theorem3}, this network is not uniformly robustly $\mathcal M$-stable. This verifies that robust set stability does not imply uniform robust set stability.

Next, the stability can be checked with respect to $I(\mathcal M)$. Using the algorithm in \cite{Guo2017Invariant}, we have
\[I(\mathcal M)=\{\ddelta_8^j\bigm| j=6,7,8\}.\]
Thus, the self-reachable states $\ddelta_8^3$ and $\ddelta_8^4$ are in $[I(\mathcal M)]^c$. By Theorem \ref{Theorem1}, this LDS is not robustly $I(\mathcal M)$-stable. This verifies that robust set stability does not necessarily imply robust stability with respect to the LRIS in the target set.

 For this example, we can also use STGs to intuitively check the robust set stability.
The STGs of the subnetworks are shown in Figure \ref{Figure1}(a), where the solid and dashed arrows represent the state transitions of the $1$st and $2$nd subnetworks, respectively. The STG $(\mathcal V,\mathcal E)$ (e.g., the union of subnetwork STGs) is shown in Figure \ref{Figure1}(b). In this figure, the red ellipse represents the set of all self-reachable states $\mathcal C_0$, which is evidently a subset of $\mathcal M$ indicating robust $\mathcal M$-stability of the network. However, parts of the states in $\mathcal C_0$ have paths to $\mathcal M^c$, indicating that the network is not uniformly robustly $\mathcal M$-stable. The blue ellipse represents the LRIS $I(\mathcal M)$ in $\mathcal M$. Evidently, it does not contain all self-reachable states, indicating that the network is not robustly $I(\mathcal M)$-stable.

 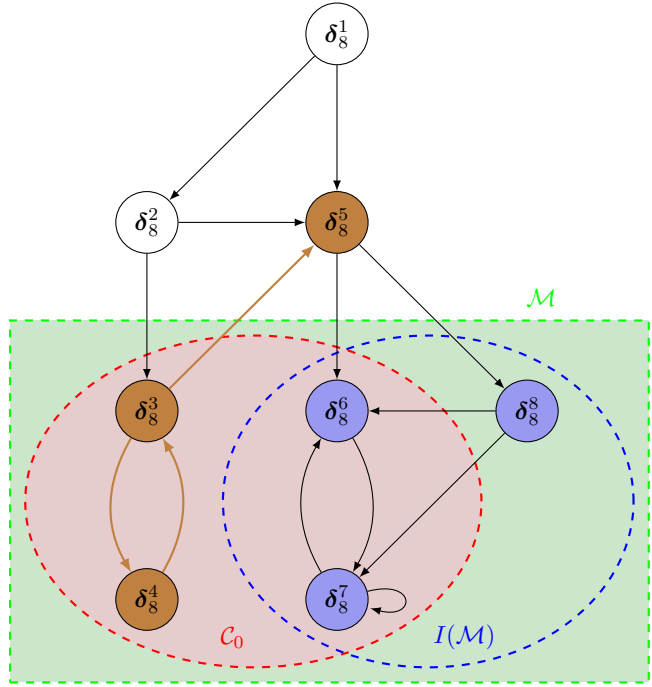
\begin{figure}[t]
    \centering

\begin{tikzpicture}[>=latex, node distance=2.5cm, on grid]
\tikzstyle{every node}=[scale=0.85]
\draw [green,fill=LGreen,dashed,thick] (-4.3,-8.6) rectangle (4.1,-3.8);
\coordinate [label=left: $\color{green}\mathcal M$ ] () at (3,-3.5);

\draw[red,dashed,thick,fill=LRed](-1.1,-6.2) ellipse (3 and 2.2 );
\coordinate [label=left: $\color{red}\mathcal C_0$ ] () at (-1.1,-8);

\draw[blue,dashed,thick](1.2,-6.2) ellipse (2.7 and 2.2 );
\coordinate [label=left: $\color{blue}I(\mathcal M)$ ] () at (2.2,-8);

\node[state] (1) {$\ddelta_8^1$};
\node[state,fill=brown] (5) [below=of 1] {$\ddelta_8^5$};
\node[state] (2) [left=of 5] {$\ddelta_8^2$};
\node[state,fill=brown] (3) [below=of 2] {$\ddelta_8^3$};
\node[state,fill=brown] (4) [below=of 3] {$\ddelta_8^4$};

\node[state,fill=LBlue] (6) [below=of 5] {$\ddelta_8^6$};
\node[state,fill=LBlue] (7) [below=of 6] {$\ddelta_8^7$};
\node[state,fill=LBlue] (8) [right=of 6] {$\ddelta_8^8$};

\path[->] (1) edge [] node {} (2);
\path[->] (2) edge [] node {} (3);
\path[->] (3) edge [bend right,color=brown,thick] node {} (4);
\path[->] (4) edge [bend right,color=brown,thick] node {} (3);
\path[->] (5) edge [] node {} (8);
\path[->] (6) edge [bend left] node {} (7);
\path[->] (7) edge [bend left] node {} (6);
\path[->] (8) edge [] node {} (7);

 \path[->] (1) edge [] node {} (5);
\path[->] (2) edge [] node {} (5);
\path[->] (3) edge [color=brown,thick] node {} (5);
\path[->] (5) edge [] node {} (6);

\path[->] (7) edge [loop right] node {} (7);
\path[->] (8) edge [] node {} (6);


\end{tikzpicture}

\caption{STG $(\mathcal V,\mathcal E)$ of LDS in Example \ref{Example1}}
\label{Figure1}
\end{figure}

\end{example}

\begin{example}\label{Example2}

 The LDS in this example is constructed from the LDS in Example \ref{Example1} by replacing the structural matrices with
  \[L_1=\ddelta_8[2,3,6,3,8,7,6,7],\quad L_2=\ddelta_8[5,5,5,7,6,7,7,6],\]
 and keeping the other parameters unchanged.
 We check the robust stability with respect to the same destination set
         \[\mathcal M=\{\ddelta_8^3,\ddelta_8^4,\ddelta_8^6,\ddelta_8^7,\ddelta_8^8\}\]
 as in Example \ref{Example1}. The union STG $(\mathcal N,\mathcal E)$ is shown in Figure \ref{Figure3}. It is easily verifiable that the set of self-reachable states is
 \[\mathcal C_0=\{\ddelta_8^6,\ddelta_8^7\}\]
 and that there are no paths from $\mathcal C_0$ to the outside of $\mathcal M$. By Theorem \ref{Theorem4}, the LDS is uniformly robustly stable with respect to $\mathcal M$. By Proposition \ref{Prop220322-1}, the LDS is also finite-time $\mathcal M$-stable with ratio one. In addition, using the algorithm in \cite{Guo2017Invariant}, the LRIS in $\mathcal M$ can be easily calculated as
 \[I(\mathcal M)=\{\ddelta_8^6,\ddelta_8^7,\ddelta_8^8\}.\]
 Clearly, $I(\mathcal M)$ contains all self-reachable states. By Proposition \ref{Proposition1}, this LDS is robustly stable with respect to $I(\mathcal M)$.

 \begin{figure}[t]
    \centering

\begin{tikzpicture}[>=latex, node distance=2.5cm, on grid]
\tikzstyle{every node}=[scale=0.85]
\draw [green,fill=LGreen,dashed,thick] (-3.2,-8.6) rectangle (4.2,-3.8);
\coordinate [label=left: $\color{green}\mathcal M$ ] () at (3,-3.5);

\draw[blue,dashed,thick](1,-6.2) ellipse (3 and 2.3 );
\coordinate [label=left: $\color{blue}{I(\mathcal M)}$ ] () at (3.2,-7);

\draw[red,dashed,thick,fill=LRed](0,-6.2) ellipse (1.7 and 1.9 );
\coordinate [label=left: $\color{red}\mathcal C_0$ ] () at (-0.7,-6.2);

\node[state] (1) {$\ddelta_8^1$};
\node[state ] (5) [below=of 1] {$\ddelta_8^5$};
\node[state] (2) [left=of 5] {$\ddelta_8^2$};
\node[state] (3) [below=of 2] {$\ddelta_8^3$};
\node[state] (4) [below=of 3] {$\ddelta_8^4$};

\node[state,fill=LBlue] (6) [below=of 5] {$\ddelta_8^6$};
\node[state,fill=LBlue] (7) [below=of 6] {$\ddelta_8^7$};
\node[state,fill=LBlue] (8) [right=of 6] {$\ddelta_8^8$};

\path[->] (1) edge [] node {} (2);
\path[->] (2) edge [] node {} (3);
\path[->] (3) edge [ ] node {} (6);
\path[->] (4) edge [ ] node {} (3);
\path[->] (4) edge [ ] node {} (7);
\path[->] (5) edge [] node {} (8);
\path[->] (6) edge [bend left] node {} (7);
\path[->] (7) edge [bend left] node {} (6);
\path[->] (8) edge [] node {} (7);

 \path[->] (1) edge [] node {} (5);
\path[->] (2) edge [] node {} (5);
\path[->] (3) edge [ ] node {} (5);
\path[->] (5) edge [] node {} (6);

\path[->] (7) edge [loop right ] node {} (7);
\path[->] (8) edge [] node {} (6);


\end{tikzpicture}

\caption{STG $(\mathcal V,\mathcal E)$ of LDS in Example \ref{Example2}}
\label{Figure3}
\end{figure}
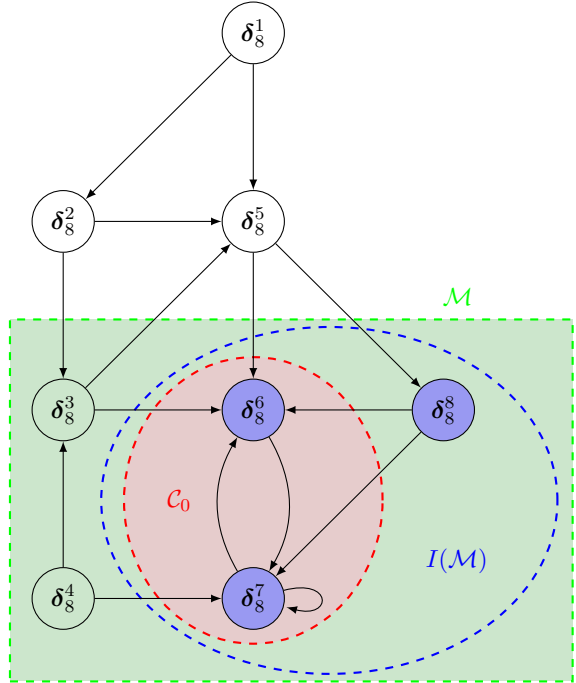

\end{example}

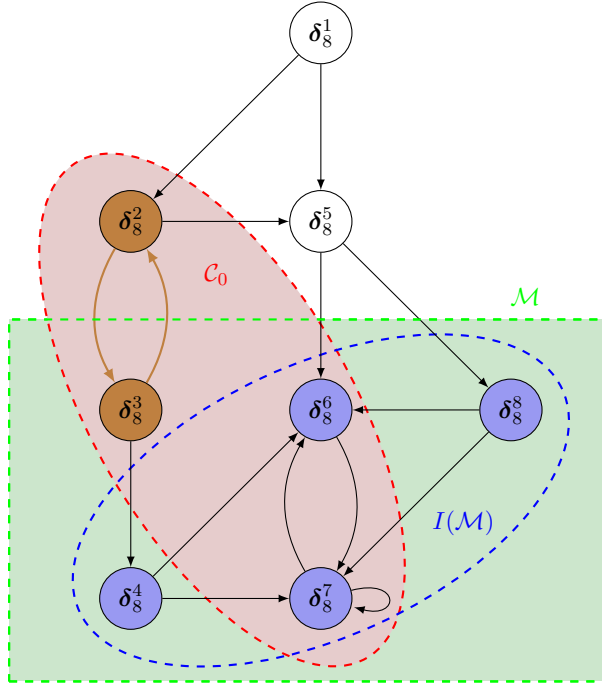
\begin{figure}

\begin{center}

\begin{tikzpicture}[>=latex, node distance=2.5cm, on grid]
\tikzstyle{every node}=[scale=0.85]
\draw [green,fill=LGreen,dashed,thick] (-4.1,-8.6) rectangle (3.8,-3.8);
\coordinate [label=left: $\color{green}\mathcal M$ ] () at (3,-3.5);

\coordinate (X) at (-1.3,-5);
\draw[rotate around={120:(X)},red,dashed,thick,fill=LRed](-1.3,-5) ellipse (3.8 and 1.7 );
\coordinate [label=left: $\color{red}\mathcal C_0$ ] () at (-1.1,-3.2);

\coordinate (X) at (0,-6.2);
\draw[rotate around={25:(X)},blue,dashed,thick](0,-6.2) ellipse (3.5 and 1.8 );
\coordinate [label=left: $\color{blue}I(\mathcal M)$ ] () at (2.4,-6.5);

\node[state] (1) {$\ddelta_8^1$};
\node[state] (5) [below=of 1] {$\ddelta_8^5$};
\node[state,fill=brown] (2) [left=of 5] {$\ddelta_8^2$};
\node[state,fill=brown] (3) [below=of 2] {$\ddelta_8^3$};
\node[state,fill=LBlue] (4) [below=of 3] {$\ddelta_8^4$};

\node[state,fill=LBlue] (6) [below=of 5] {$\ddelta_8^6$};
\node[state,fill=LBlue] (7) [below=of 6] {$\ddelta_8^7$};
\node[state,fill=LBlue] (8) [right=of 6] {$\ddelta_8^8$};

\path[->] (1) edge [] node {} (2);
\path[->] (2) edge [bend right,color=brown,thick] node {} (3);
\path[->] (3) edge [] node {} (4);
\path[->] (4) edge [] node {} (7);
\path[->] (4) edge [] node {} (6);
\path[->] (5) edge [] node {} (8);
\path[->] (6) edge [bend left] node {} (7);
\path[->] (7) edge [bend left] node {} (6);
\path[->] (8) edge [] node {} (7);

 \path[->] (1) edge [] node {} (5);
\path[->] (2) edge [] node {} (5);
\path[->] (3) edge [bend right, color=brown,thick] node {} (2);
\path[->] (5) edge [] node {} (6);

\path[->] (7) edge [loop right] node {} (7);
\path[->] (8) edge [] node {} (6);


\draw [green,dashed,thick] (-4.1,-8.6) rectangle (3.8,-3.8);

\end{tikzpicture}

\end{center}

\caption{STG $(\mathcal V,\mathcal E)$ of LDS in Example \ref{Example3}}
\label{Figure4}

\end{figure}

\begin{example}\label{Example3}
     Consider an LDS (\ref{LDS-AlgebraicForm1}) with $N = 8$, $M = 2$, and
    \[L_1=\ddelta_8[2,3,2,6,6,7,6,6],\quad L_2=\ddelta_8[5,5,3,7,8,7,7,7].\]
    We assume that the target subset is
   \[\mathcal M=\{\ddelta_8^3,\ddelta_8^4,\ddelta_8^6,\ddelta_8^7,\ddelta_8^8\}. \]
   A simple calculation yields
   \[I(\mathcal M)=\{\ddelta_8^4, \ddelta_8^6,\ddelta_8^7,\ddelta_8^8\}.\]
   The STG $(\mathcal V,\mathcal E)$ is given in Figure \ref{Figure4}, where $I(\mathcal M)$ contains the states in the blue ellipse. It is easily verified  that every vertex STG $(\mathcal V,\mathcal E)$ has a path to $I(\mathcal M)$. Thus, by Theorem \ref{theorem6}, the LDS is asymptotically $\mathcal M$-stable with ratio one. However,  the set of all self-reachable states, which contains the states in the red ellipse, is
 \[\mathcal C_0=\{\ddelta_8^2,\ddelta_8^3,\ddelta_8^6,\ddelta_8^7\}.\]
 Thus, $\mathcal M^c$ contains a self-reachable state $\ddelta_8^2$. By Proposition \ref{Proposition1}, this LDS is not robustly $\mathcal M$-stable. This example verifies that asymptotical set stability with ratio one does not imply robust set stability.

\end{example}

\section{Conclusion}\label{SectionConcluding}

Different definitions of robust stability for LDSs with respect to uncertain switching are presented herein, including robust set stability, uniform robust set stability, asymptotical set stability with ratio one, and finite-time set stability with ratio one.  It is proved herein that, although robust set stability and uniform robust set stability become equivalent if the LDS is uncertainty-free or the destination set is a singleton, they are essentially different. Specifically, uniform robust set stability implies robust set stability; however, the inverse claim is untrue. In addition, robust set stability implies asymptotical set stability with ratio one; however, the inverse claim is not true. Especially, the concept of set stability with ratio one bridges robust stability and stability of LDSs under random switching. Asymptotical and finite-time set stability with ratio one are proved to be equivalent to asymptotical set stability in distribution and finite-time set stability with probability one of LDSs under i.i.d. random switching.

 Necessary and sufficient conditions for different stabilities, which are easily verifiable, have been obtained.  Specifically, we proved the following:
 \begin{enumerate}
   \item An LDS is robustly set stable if and only if the states outside the destination set are all self-unreachable.
   \item An LDS is uniformly robustly set stable, or finite-time set stable with ratio one, if and only if any state outside the destination set is unreachable from any self-reachable state.
   \item An LDS is asymptotically set stable if and only if the LRIS in the destination set is reachable from any state.
 \end{enumerate}

 The largest invariant subset-based analysis is an important technique to obtain stability criteria for both deterministic and probabilistic Boolean networks in the literature. This study indicates that caution should be exercised when this technique is applied to robust stability analysis for LDSs under uncertain switching. It is proved herein that, for uniform robust set stability and asymptotical set stability with ratio one, the set stability is equivalent to the stability with respect to the LRIS in the destination set. However, robust set stability is not equivalent to robust set stability with respect to the LRIS in the destination set. This finding corrects an incorrect result in a previous study \cite{Guo2017Invariant} concerning robust set stability.

 A problem that needs to be addressed in the future is the corresponding robust set stabilization. Through the LRIS-based results obtained in this study, the technique of state space decomposition for feedback set stabilization of an LDS proposed in \cite{Guo2015Set} is potentially applicable for uniform robust feedback set stabilization of LDSs under uncertain switching. However, this technique is invalid for robust feedback set stabilization because there are no corresponding LRIS-based criteria. Therefore, new methods need to be developed to address this problem.

	\bibliographystyle{IEEEtranS}
	\bibliography{BooleanRef}

%
%
%


\end{document}